\documentstyle[amssymb,aps]{revtex}

\large

\begin{document}
\title{Viability of competing field theories for the driven lattice gas.}
\author{B. Schmittmann$^{1}$, H.K. Janssen$^{2}$, U.C. T\"{a}uber$^{1}$, R.K.P. Zia$%
^{1}$, K.-t. Leung$^{3}$, J.L. Cardy$^{4}$}
\address{$^1$Center for Stochastic Processes in Science and Engineering and 
Department of Physics, \\
Virginia Tech, Blacksburg, VA 24061-0435 USA;\\
$^2$Institut f\"{u}r Theoretische Physik III,
Heinrich-Heine-Universit\"{a}t, D-40225 D\"{u}sseldorf, Germany;\\
$^3$Institute of Physics, Academia Sinica, Taipei, Taiwan 11529, ROC;\\
$^{4}$Theoretical Physics, 1 Keble Road, Oxford OX1 3NP, UK.}
\date{\today }
\maketitle

\begin{abstract}
It has recently been suggested that the driven lattice gas should be
described by a novel field theory in the limit of infinite drive. We review
the original and the new field theory, invoking several well-documented key
features of the microscopics. Since the new field theory fails to reproduce
these characteristics, we argue that it cannot serve as a viable description
of the driven lattice gas. Recent results, for the critical exponents
associated with this theory, are re-analyzed and shown to be incorrect.
\end{abstract}




The critical behavior of the driven lattice gas (DLG) \cite{KLS} has been
the subject of some debate, ever since the first Monte Carlo simulations 
\cite{earlyMC} and field theoretic predictions \cite{JS,LC} were found to
give differing values for the order parameter exponent $\beta $. This
discrepancy has led to developments in different directions: some
researchers \cite{KTL,JSW} have modified the simulation data analysis,
invoking anisotropic finite size scaling \cite{AFSS}, while others \cite
{PRE1,JSP} have suggested that the original field theory might be deficient
in the limit of infinite drive, proposing \cite{PRE1,JSP} and analyzing \cite
{PRE2} an alternate coarse-grained theory instead. In this communication, we
review both the original \cite{JS,LC} and the alternate \cite{PRE1,JSP,PRE2}
field theory, in the light of Monte Carlo simulation data. We first document
that the alternate theory is {\em not} a coarse-grained description of the
driven lattice gas, since it fails to exhibit several well-established
properties of the microscopic model. In a second step, we re-analyze the
proposed theory, assuming that it might describe some other, yet to be
determined, microscopics. We show that the renormalization group analysis of
Ref. \cite{PRE2} is {\em seriously flawed}, resulting in incorrect exponents
and a proliferation of uncontrolled infrared singularities.

We begin with a brief summary of the background. Microscopically, the DLG is
a simple ferromagnetic Ising lattice gas, half-filled and coupled to a heat
bath at temperature $T$, in which particles jump to empty nearest-neighbor
sites subject to the usual Ising energetics and a uniform driving force $E$
acting along a particular lattice direction. Thus, the effect of $E$ is
identical to adding a locally linear potential. Clearly, $E=0$ corresponds
to the equilibrium Ising model. On the other hand, even $E=\infty $ can be
realized if Metropolis rates are used: Simply accept/forbid all
forward/backwards jumps. Since large values of $E$ accentuate the
nonequilibrium features of this system, most simulations have been performed
at $E\gtrsim 50$, in units of the Ising coupling constant.

The driven lattice gas and many of its variants have attracted considerable
attention since they evolve into simple nonequilibrium steady states
displaying a wealth of counterintuitive characteristics \cite{SZ}. Two of
its most remarkable features are ({\em i}) the discontinuity singularity of
the structure factor $S({\bf k})$ \cite{KLS,sf}, which is intimately
connected to an $r^{-d}$ decay (in $d$ dimensions) of the two-point 
correlations \cite{ZWLV,r-d}%
, and ({\em ii}) the emergence of nontrivial three-point correlations \cite
{3pf} in the disordered phase, corresponding to the violation of the Ising
symmetry by $E$ (which drives particles and holes in opposite directions).
Such dramatically ``non-Ising'' characteristics are easily observed in Monte
Carlo simulations, at intermediate and large driving fields. They are also
confirmed in a high-temperature series expansion, derived directly from the
microscopic dynamics \cite{ZWLV,hts}.

These observations from Monte Carlo simulations play a crucial role in
identifying the correct field theory. A basic tenet in the study of critical
phenomena is that a microscopic model and its coarse-grained field theory
should possess the same symmetries, if they are to belong into the same
universality class. For the driven lattice gas, the data on the structure
factor indicate that the theory is highly {\em anisotropic}. Moreover, the
detailed behavior of the discontinuity singularity, upon approaching the
origin in wave vector space from different directions, informs us {\em %
precisely how} the familiar Ornstein-Zernike form is modified. Generically,
we find \cite{SZ} that 
\begin{equation}
R\equiv \frac{\lim_{|{\bf k}_{\bot }|\rightarrow 0}S({\bf k}_{\bot
},k_{\Vert }=0)}{\lim_{k_{\Vert }\rightarrow 0}S({\bf k}_{\bot }=0,k_{\Vert
})}>1  \label{SF MC}
\end{equation}
above criticality, and $R\rightarrow \infty $ upon approaching $T_c$. The
subscripts distinguish the parallel ($\Vert $) and transverse ($\bot $)
subspaces, with respect to the drive direction. Just as significantly, the
non-vanishing three-point functions demonstrate that the usual ``up-down''
symmetry of the Ising model is broken.

These key features of the microscopics must be reflected in any viable
continuum theory for the driven lattice gas. We first consider the original
field theory\cite{JS,LC}. It is based on a Langevin equation, in continuous
space and time, which describes the stochastic evolution of the local
particle density $\rho ({\bf x},t)$. In terms of $\phi \equiv 2\rho -1$, the
equation reads: 
\begin{equation}
\partial _{t}\phi =\lambda \left\{ \left( \tau _{\bot }-\nabla _{\perp
}^{2}\right) \nabla _{\perp }^{2}\phi +\tau _{\Vert }\nabla _{\Vert
}^{2}\phi +{\cal E}\nabla _{\Vert }\phi ^{2}+\frac{g}{3!}\nabla _{\perp
}^{2}\phi ^{3}\right\} -{\bf \nabla \xi \ .}  \label{LE1}
\end{equation}
The Langevin noise term reflects the fast degrees of freedom: 
\begin{eqnarray*}
\left\langle {\bf \xi (x,t)}\right\rangle &=&0 \\
\left\langle {\bf \nabla \xi }({\bf x},t){\bf \nabla }^{\prime }{\bf \xi }(%
{\bf x}^{\prime },t^{\prime })\right\rangle &=&-2\left( n_{\bot }\nabla
_{\perp }^{2}+n_{\Vert }\nabla _{\Vert }^{2}\right) \delta ({\bf x}-{\bf x}%
^{\prime })\delta (t-t^{\prime })\ .
\end{eqnarray*}
We emphasize that ({\em i}) all coefficients are strictly positive {\em %
except }possibly $\tau _{\bot }$ and/or $\tau _{\Vert }$ which control
criticality (see below) and ({\em ii}) independent from one another (i.e.,
not related by symmetry). The parameter $\lambda $ sets the time scale.

This theory contains two closely linked key ingredients: First, there is a
driving term, ${\cal E}\nabla _{\Vert }\phi ^{2}$, where ${\cal E}$ denotes
the coarse-grained drive (a naive continuum limit gives ${\cal E}\propto
\tanh (E/T)$). This term is {\em required} to break the Ising ``up-down'' ($%
\phi \rightarrow -\phi $) symmetry. Second, the theory is highly
anisotropic, with two {\em different} diffusion coefficients $\tau _{\Vert }$
and $\tau _{\perp }$. In particular, it predicts an equal-time structure
factor, 
\begin{equation}
S({\bf k})=\frac{n_{\bot }k_{\bot }^{2}+n_{\Vert }k_{\Vert }^{2}}{\tau
_{\bot }k_{\bot }^{2}+\tau _{\Vert }k_{\Vert }^{2}+O(k^{4})}\text{ }
\label{SF1}
\end{equation}
in the disordered phase. This $S$ generates a discontinuity singularity $%
R=(n_{\bot }\tau _{\parallel })/(n_{\Vert }\tau _{\bot })$. To ensure that
the {\em observed} behavior is faithfully reproduced, we demand $n_{\bot
}\tau _{\Vert }$ $>n_{\Vert }\tau _{\perp }$ in the disordered phase.
Moreover, criticality {\em must} be marked by $\tau _{\perp }=0$ at {\em %
positive} $\tau _{\Vert }$ if the divergence of $S$ is to be captured
correctly. To summarize, the two key features of the original Langevin
equation are unambiguously supported by the Monte Carlo data for the
microscopic model.

We comment briefly on the issue of {\em finite }versus {\em infinite\ }%
fields. In all Monte Carlo simulations, the current is observed to saturate
as $E$ increases. This saturation is reflected by ${\cal E}\propto \tanh
(E/T)\rightarrow 1$ in the original field theory. Therefore, this theory
holds equally well for any nonzero value of the microscopic drive.
Furthermore, simulations using {\em Metropolis rates} with $E=50$, $100$ and 
$\infty $ have been performed. The results are (statistically)
indistinguishable! Such sensible behavior is entirely consistent with this
theory.

The discrepancies arise when critical exponents are measured, specifically
the order parameter exponent $\beta $, and compared to field theoretic
predictions. The original field theory, due to the vanishing of $\tau
_{\perp }$ at positive $\tau _{\parallel }$, naturally leads to anisotropic
scaling of wave vectors: $k_{\Vert }\sim k_{\perp }^{1+\Delta }$ in the
critical region, with a nontrivial anisotropy exponent $\Delta $. Three
important consequences are that, first, the upper critical dimension $d_{c}$
is shifted to $5$, and second, the theory predicts $\Delta =$ $1+(5-d)/3$
and $\beta =1/2$ {\em exactly}, i.e., to all orders in perturbation theory.
The values obtained by simulations differ, depending on the method used to
analyze the data. If a careful anisotropic finite size analysis \cite{AFSS}
is used, based on system sizes consistent with the field-theoretic scaling,
i.e., $L_{\Vert }/L_{\perp }^{1+\Delta }=const$, the field-theoretic
exponents result in good data collapse for a number of different observables 
\cite{KTL,JSW}. However, data for isotropic systems, $L_{\Vert }/L_{\perp
}=const$, appear to indicate an order parameter exponent around $0.23$ \cite
{earlyMC}. Since most of the data were taken at very large fields, some
authors \cite{PRE1,JSP} have suggested that the origin of the discrepancies
does not reside in the data analysis. Instead, they argue that the standard
field theory does not capture the $E\rightarrow \infty $ limit correctly and
propose an alternate theory. It is based on the Langevin equation: 
\begin{equation}
\partial _{t}\phi =\lambda \left\{ \left( \tau _{\bot }-\nabla _{\perp
}^{2}\right) \nabla _{\perp }^{2}\phi -\nabla _{\Vert }^{2}\nabla _{\perp
}^{2}\phi +\frac{g}{3!}\nabla _{\perp }^{2}\phi ^{3}\right\} -{\bf \nabla
\xi \ .}  \label{LE2}
\end{equation}
With minor renamings of parameters \cite{not}, this is Eq. (1) of Ref. \cite
{PRE2}. The vanishing of $\tau _{\bot }$ marks the critical point. The noise
satisfies (Eq. (2) of Ref. \cite{PRE2}): 
\begin{eqnarray*}
\left\langle {\bf \xi (x,}t{\bf )}\right\rangle &=&0 \\
\left\langle {\bf \nabla \xi }({\bf x},t){\bf \nabla }^{\prime }{\bf \xi }(%
{\bf x}^{\prime },t^{\prime })\right\rangle &=&-2\lambda \left( \nabla
_{\perp }^{2}+\frac{1}{2}\nabla _{\Vert }^{2}\right) \delta ({\bf x}-{\bf x}%
^{\prime })\delta (t-t^{\prime })\ .
\end{eqnarray*}
Two key terms appearing in the original field theory are absent in this one,
namely,

\begin{itemize}
\item  the driving term $\nabla _{\Vert }\phi ^{2}$, and

\item  a diffusion term $\tau _{\Vert }\nabla _{\Vert }^{2}\phi $ for the
parallel direction.
\end{itemize}

Since the driving term is absent, the alternate field theory obeys the Ising
``up-down'' ($\phi \rightarrow -\phi $) symmetry. Thus, three-point
functions are identically {\em zero} in this theory, for all $T\geq T_{c}$.
This prediction is in serious disagreement with existing Monte Carlo data!
While one may argue that a field theory need not reproduce all of the
microscopic detail of the underlying lattice model, one should be very
cautious before endowing it with a {\em higher symmetry}: This is only
justified if a high-symmetry fixed point exists and can be shown, via an
explicit renormalization group calculation, to be {\em stable} against
perturbations by symmetry-breaking operators. Neither is the case here.

The absence of the parallel diffusion term also has serious consequences.
Eq. (\ref{LE2}) generates a steady-state structure factor: 
\begin{equation}
S({\bf k})=\frac{k_{\bot }^2+\frac 12k_{\Vert }^2}{k_{\bot }^2\left(
k^2+\tau _{\bot }\right) }\text{ }  \label{SF2}
\end{equation}
which ought to be a good approximation at high temperatures. Yet, for $%
k_{\Vert }\neq 0$ it predicts a {\em divergence} along the {\em whole} $%
k_{\bot }=0$ line, at {\em any} $T>T_c$. This stands in glaring contrast to
the Monte Carlo results for the disordered phase, where {\em all} measured
structure factors are found to be finite {\em everywhere} in $k$-space.

Since Eq. (\ref{LE2}) fails to reproduce the most basic properties of the
microscopic model, we conclude that it is not a viable field theory for the
driven lattice gas. It may, however, describe some as yet unknown
microscopics. Therefore, we now proceed to analyze the field theory, defined
by Eq. (\ref{LE2}), in its own right.

Following Ref. \cite{PRE2}, we recast Eq. (\ref{LE2}) as a dynamic
functional \cite{df}: 
\begin{equation}
{\cal L}[\tilde{\phi},\phi ]=\int d^{d}xdt\left\{ \tilde{\phi}\left[
\partial _{t}\phi +\lambda \left( \nabla _{\Vert }^{2}\nabla _{\perp
}^{2}+(\nabla _{\perp }^{2})^{2}-\tau _{\bot }\nabla _{\perp }^{2}\right)
\phi -\lambda \frac{g}{3!}\nabla _{\perp }^{2}\phi ^{3}\right] -\lambda 
\tilde{\phi}\left( \nabla _{\perp }^{2}+\frac{1}{2}\nabla _{\Vert
}^{2}\right) \tilde{\phi}\right\}  \label{DF}
\end{equation}
We first note that Eq. (\ref{DF}) describes a theory with a four-point
coupling $\tilde{\phi}\nabla _{\perp }^{2}\phi ^{3}$ and anisotropic free
propagators as given in Ref. \cite{PRE2}. Therefore, the combinatorics of
this theory is identical to that of Model B, which reduces to $\phi ^{4}$%
-theory in the static limit. For such theories, it is well known \cite{DJA}
that the one-loop result for the exponent $\nu $ (denoted $\nu _{\perp }$ in
Ref. \cite{PRE2}) is determined by {\em combinatorics alone}, i.e., the
explicit expressions for the Feynman integrals are not required. This is
most easily seen by calculating in the critical theory, where $\tau _{\bot
}=0$, with insertions of $\lambda \tilde{\phi}\nabla _{\perp }^{2}\phi $. We
denote one-point irreducible vertex functions with $\tilde{n}$ ($n$)
external $\tilde{\phi}$ ($\phi $) legs and $m$ insertions by $\Gamma _{%
\tilde{n}n}^{(m)}$. At one-loop order, there are two primitively divergent
vertex functions, namely $\Gamma _{11}^{(1)}$ and $\Gamma _{13}^{(0)}$. Both
of these consist of a zero-loop term and a one-loop contribution. Each
one-loop contribution consists of a combinatoric factor, the appropriate
powers of the coupling constant and the external momentum, and a loop
integral. The {\em key }simplification here is that the loop {\em integrals}
for $\Gamma _{11}^{(1)}$ and $\Gamma _{13}^{(0)}$ are {\em identical},
independent of the detailed forms of the free propagators. Thus, the two
one-loop contributions differ only by a simple factor which is purely
combinatoric in origin. As a result, one obtains to first order in $\epsilon
\equiv d_{c}-d$, for {\em all} of these theories: 
\begin{equation}
\nu =\frac{1}{2}+\frac{\epsilon }{12}+O(\epsilon ^{2})  \label{nu}
\end{equation}
Since the authors of Ref. \cite{PRE2} have chosen to calculate at finite $%
\tau _{\bot }$, let us illustrate how this result emerges in their case. No
insertions are needed here so the upper index of $\Gamma _{\tilde{n}n}^{(m)}$
can be dropped. Keeping track of coupling constants and signs, and taking
care of the $T_{c}$ shift, we can write the two bare vertex functions $%
\Gamma _{11}$ and $\Gamma _{13}$ in the form 
\begin{eqnarray}
\Gamma _{11} &=&i\omega +\lambda k_{\perp }^{2}k^{2}+\lambda k_{\perp
}^{2}\tau _{\bot }\left[ 1+\frac{1}{2}gI_{1}\right]  \nonumber \\
\Gamma _{13} &=&\lambda gk_{\perp }^{2}\left[ 1-\frac{3}{2}gI_{2}\right]
\label{gams}
\end{eqnarray}
Here, the factors $1/2$ and $-3/2$ arise from combinatorics while the
integrals $I_{1}$ and $I_{2}$ are easily computed in dimensional
regularization, resulting in 
\begin{equation}
I_{1}=\frac{3}{(4\pi )^{2}\epsilon }\left[ 1+O(\epsilon )\right] \;\;\text{%
and\ \ }I_{2}=\frac{3}{(4\pi )^{2}\epsilon }\left[ 1+O(\epsilon )\right]
\end{equation}
We notice immediately that the simple $\epsilon $-poles of $I_{1}$ and $%
I_{2} $ are {\em identical}. Thus, their numerical prefactor can be absorbed
into the definition of the coupling constant, leaving us with one-loop
corrections to $\nu $ that are purely combinatoric in origin. Completing the
calculation at finite $\tau _{\bot }$, this provides the key to Eq. (\ref{nu}%
). Only at {\em two-loop} order do the detailed forms of the free
propagators come into play. Then, of course, exponents are also no longer
determined by combinatorics alone.

In Ref. \cite{PRE2}, the exponent $\nu $ is quoted as $(1+\epsilon /4)/2$,
indicating the presence of a computational error. More seriously, however,
there are deeper flaws in this theory. Recall that the steady-state
structure factor, Eq. (\ref{SF2}), diverges along the whole $k_{\bot }=0$
line, even for $\tau _{\bot }>0$. As a result, the theory is plagued by {\em %
infrared} singularities, which are entirely {\em unrelated} to criticality,
and unrenormalizable divergences. We note, for completeness, that one can,
of course, regularize such singularities by {\em re-introducing} the diffusion
term $\tau _{\Vert }\nabla _{\Vert }^{2}\phi $ into Eq. (\ref{LE2}). Then,
however, one should also reconsider the two {\em fourth-order} derivative
terms. At two-loop order, these will acquire {\em different} primitive
divergences, so that an additional coupling constant $\rho ^{2}$ is
required, appearing in Eq. (\ref{LE2}) as $\rho ^{2}(\nabla _{\perp
}^{2})^{2}\phi +\nabla _{\Vert }^{2}\nabla _{\perp }^{2}\phi $.

To summarize, we have shown that the field theory proposed by Garrido, de
los Santos and Mu\~{n}oz \cite{PRE1,JSP} fails to reproduce the key features
of the driven lattice gas. Predicting {\em infinite} structure factors and 
{\em zero} three point correlations (for {\em all} temperatures above
criticality), it cannot be a viable continuum model for the latter.
Accepting it as a representation of some other, as yet undetermined,
microscopic model, we carry out a standard analysis. First, we find that the
one-loop calculation of Ref. \cite{PRE2} is incorrect. Second, beyond
one-loop order, uncontrolled infrared singularities proliferate, rendering
the field theory unrenormalizable. In contrast, the original field theory 
\cite{JS,LC} is consistent with the fundamental symmetries of the driven
lattice gas, for any value of the drive. Its predictions for 2- and 3-point
functions in the disordered phase are in good agreement with simulation
results. Based on the phenomenology of $S({\bf k})$ near criticality, it
plumbs the consequences of a highly anisotropic scaling limit, $k_{\Vert
}\sim k_{\bot }^{1+\Delta }\rightarrow 0$. To test its predictions against
Monte Carlo simulations, this limit should be respected in the choice of
system sizes, i.e., $L_{\Vert }\sim L_{\bot }^{1+\Delta }$. If, instead,
simulations and finite-size analysis are performed with disregard for such
strong anisotropies, complications from extraneous scaling variables \cite
{KTL} or inconsistencies \cite{KTL-JSW} can be expected. In a more exotic
scenario, such simulations may indicate a new type of low temperature phase,
quite distinct from the ordinary Ising-like one.

This work is supported in part by the National Science Foundation through
the Division of Materials Research. \newline


\begin{references}
\bibitem{not}  In Ref. \cite{PRE2}, $\lambda $ is denoted by $e_o/2$ and $%
\tau _{\bot }$ is just $\tau $. Our renamings make comparison of Eqs. (\ref
{LE1}) and (\ref{LE2}) easier.
\end{references}
\end{document}